\documentclass[a4paper,aip,apl,amsmath,amssymb,reprint,superscriptaddress]{revtex4-1}
\pdfoutput=1
\usepackage{graphicx}
\usepackage{amsmath}
\usepackage{dcolumn}
\usepackage{bm}
\usepackage{bbold}
\usepackage{color}
\usepackage{amsfonts}
\usepackage{amssymb}
\usepackage{mathrsfs}
\usepackage{tabularx}
\usepackage{braket}
\usepackage{mathtools}
\usepackage{soul}			

\usepackage{caption}
\usepackage{subcaption}

\newcommand{\deriv}[2][t]{\frac{\mathrm{d}#2}{\mathrm{d}#1}}

\usepackage[normalem]{ulem} 

\begin{document} 
\title{Broadening Frequency Range of a Ferromagnetic Axion Haloscope with Strongly Coupled Cavity-Magnon Polaritons}

\author{Graeme Flower}
\email{21302579@student.uwa.edu.au}
\affiliation{ARC Centre of Excellence for Engineered Quantum Systems, Department of Physics, University of Western Australia, 35 Stirling Highway, Crawley WA 6009, Australia}

\author{Jeremy Bourhill}
\affiliation{ARC Centre of Excellence for Engineered Quantum Systems, Department of Physics, University of Western Australia, 35 Stirling Highway, Crawley WA 6009, Australia}

\author{Maxim Goryachev}
\affiliation{ARC Centre of Excellence for Engineered Quantum Systems, Department of Physics, University of Western Australia, 35 Stirling Highway, Crawley WA 6009, Australia}

\author{Michael E. Tobar}
\email{michael.tobar@uwa.edu.au}
\affiliation{ARC Centre of Excellence for Engineered Quantum Systems, Department of Physics, University of Western Australia, 35 Stirling Highway, Crawley WA 6009, Australia}

\date{\today}


\begin{abstract}

With the axion being a prime candidate for dark matter, there has been some recent interest in direct detection through a so called `Ferromagnetic haloscope.' Such devices exploit the coupling between axions and electrons in the form of collective spin excitations of magnetic materials with the readout through a microwave cavity. Here, we present a new, general, theoretical treatment of such experiments in a Hamiltonian formulation for strongly coupled magnons and photons, which hybridise as cavity-magnon polaritons. Such strongly coupled systems have an extended measurable dispersive regime. Thus, we extend the analysis and operation of such experiments into the dispersive regime, which allows any ferromagnetic haloscope to achieve improved bandwidth with respect to the axion mass parameter space. This experiment was implemented in a cryogenic setup, and initial search results are presented setting laboratory limits on the axion-electron coupling strength of  $g_{aee}>3.7\times10^{-9}$ in the range $33.79~\mu$eV$< m_a<33.94~\mu$eV with $95\%$ confidence. The potential bandwidth of the Ferromagnetic haloscope was calculated to be in two bands, the first of about $1$~GHz around $8.24$~GHz (or $4.1~\mu$eV mass range around $34.1~\mu$eV) and the second of about $1.6$~GHz around $10$~GHz  ($6.6~\mu$eV mass range around $41.4~\mu$eV). Frequency tuning may also be easily achieved via an external magnetic field which changes the ferromagnetic resonant frequency with respect to the cavity frequency. The requirements necessary for future improvements to reach the DFSZ axion model band are discussed in the paper.

\end{abstract}

\maketitle

\section*{Introduction}
Dark matter continues to be one of the major unsolved problems in physics. Evidence of its existence is abundant but its identity remains a mystery. Many physicists believe it consists of particles that come from extensions to the current Standard Model (SM). The axion is one such particle that comes from an SM extension originally proposed to solve a problem in Quantum Chromodynamics known as the strong charge parity problem \cite{CPaxions}. It has since been shown to make an excellent candidate for dark matter, potentially able to account for all of dark matter \cite{darkaxions,Jaeckel:2010aa}. A key parameter in axion theory is the Pecci-Quinn symmetry breaking scale ($f_a$) which determines the mass and coupling strengths to ordinary matter. The most known model parameter bands are KSVZ (Kim-Shifman-Vainshtein-Zakharov)\cite{PhysRevLett.43.103,Shifman:1980aa} and DFSZ (Dine-Fischler-Srednicki-Zhitnitsky)\cite{DFSZaxion2,Zhitnitskij:1980aa}. Despite existence of several theoretical proposals\cite{Ballesteros:2017aa,PhysRevD.93.035001}, this parameter is unknown with limited bounds on it. As such there is a vast portion of the parameter space of axion mass and couplings to search making the first observation of axion or axion like particles very challenging\cite{Irastorza:2018aa}.  \\

The most successful implementation of axion direct detection experiments\cite{Graham:2015aa} are, so called, axion haloscopes, such as the Axion Dark Matter eXperiment (ADMX)\cite{admx,Collaboration:2018aa}, HAYSTAC\cite{haystac}, CULTASK\cite{CULST}, ORGAN\cite{organ}, ACTION\cite{PhysRevD.96.061102} and multi-cell cavity approaches\cite{JEONG2018412,JEONG201833}. These experiments focus on photon coupling to galactic axion halo probing the hypothesis that axions constitute the bulk of dark matter in the Milky Way as opposed to helioscopes\cite{Ribas:2015aa,Dafni:2016aa} focusing on axion and Axion Like Particles sourced from the Sun.  Mentioned above haloscopes exploit the inverse Primakoff effect\cite{sikivie} and strong DC magnetic fields as virtual photons to detect axions at RF and microwave frequencies putting limits on axion-photon coupling parameter $g_{a\gamma\gamma}$. The same coupling is investigated using a slightly modified approach based on dielectric materials in experiments such as MADMAX\cite{Group:2017aa}. On the other hand, ferromagnetic haloscopes, considered in this work, probe axion coupling to electron spin, coupling parameter $g_{aee}$. Such experiments probe the interaction between axions in the dark matter halo and collective spin excitations in ferromagnetic materials, quantized version of which is widely known as a magnon. This resultant signal is then readout through coupling the magnetic mode to a photon mode by placing the magnetic sample inside a microwave cavity. Similar in spirit to ferromagnetic haloscope, searches with nuclear spins, such as CASPER\cite{Garcon:2018aa}, constitute another class of galactic haloscope targeting another coupling parameter $g_{aNN}$. \\

There has been some recent interest in this scheme with the QUaerere AXion (QUAX) experiment\cite{Crescini2018,quaxprop,QUAXOriginal}. Currently, though, the QUAX experiment is largely static in frequency due to it operating in the, so called, fully hybridised regime, and so has limited capability in scanning axion mass, a critical feature for an axion haloscope. In the recent development of this setup\cite{Crescini2018}, two methods to make the experiment frequency tunable are proposed. The first approach is to tune both the cavity and magnon mode equally to move the hybrid frequencies. This approach, while achievable in principle, adds significant complexity to the experiment. The second tuning method for a ferromagnetic haloscope is to detune the magnetic resonance from the cavity resonance. Though, this is practical and could be easily implemented, no method to determine how the experiment sensitivity changes in such a regime is provided. In the present work, we cover this lack of knowledge about this method by considering a new approach to determining the dynamics of such the cavity-magnon system.

\section{Detection Principles}
\subsection{Axion Electron Interaction}

To understand working principles of magnetic axion haloscopes, one needs to consider interaction between axions and an individual electron. The starting point here is the following Lagrangian for the DSVZ axion coupled to fermionic field:
\begin{equation}
\mathcal{L} = \bar{\psi}(x)(i\hbar\gamma^\mu\partial_\mu-mc)\psi(x) - ig_{aee}a(x)\bar{\psi}(x)\gamma_5\psi(x),
\end{equation}
where $\psi(x)$ is the spinor field of the fermion, $m$ is its mass, $a(x)$ is the axion field, $\gamma^\mu$ are the Dirac matrices and $g_{aee}$ is the axion to electron coupling strength. Based on this Lagrangian and ignoring the, so called, parity violating term one can derive a Hamiltonian operator for an electron in the non-relativistic limit:
\begin{equation}
\hat{H} = -\frac{\hbar^2}{2m_e}\nabla^2 - \frac{g_{aee}\hbar^2}{2m_e}\hat{\sigma} \cdot \nabla \mathbf{a},
\label{CP001}
\end{equation}
where $e$ is the electron charge, $\hat{\sigma}$ is the Pauli matrices vector. This Hamiltonian can be compared to the one for an electron in a magnetic field $\mathbf{B}$:
\begin{equation}
\hat{H} = -\frac{\hbar^2}{2m_e}\nabla^2 - \frac{e\hbar^2}{m_e}\hat{\sigma} \cdot \mathbf{B}.
\label{CP002}
\end{equation}
The similarity between Hamiltonians (\ref{CP001}) and (\ref{CP002}) allows to introduce the pseudo-magnetic field\cite{origmagnonaxion,pseudoB1,pseudoB2,CPaxions,Crescini2018}:
\begin{equation}
\mathbf{B}_{aee} = \frac{g_{aee}}{2e}\nabla \mathbf{a}.
\label{CP003}
\end{equation}
Thus, the axion effects on electrons could be regarded as an effect of an anomalous oscillating magnetic field. Parameters of these oscillations as well as its direction are considered later in Section \ref{windS}. Thus far only the DFSZ axion has been considered. This is due to the suppression of the KSVZ axion to ordinary quark and lepton coupling at the tree level \cite{KSVZaxion1,KSVZaxion2}. The coupling constant, $g_{aee}$, for the DFSZ axion, however, is given by:
\begin{equation}
g_{aee} = \frac{m_e}{3f_a}\cos^2(\beta),
\end{equation}
where $\cot(\beta)$ is the ratio of two Higgs vacuum expectation values of this model \cite{DFSZaxion1,DFSZaxion2}. For the following work, it is assumed $\cos(\beta)=1$.

\subsection{Ferromagnetic Haloscopes}
\label{ferroHalo}

In order to maximise the axion induced signal in a realistic detector, it is important to shift from a single spin system to a system with a very high spin number. This number of spins is maximised in ferromagnetic crystals, particularly Yttrium Iron Garnet (YIG), that have been proposed as the core sensing element in previous proposals\cite{origmagnonaxion,quaxprop,Crescini2018}. In addition to this, high quality YIG is characterised by relatively narrow linewidths that makes it a popular choice for research in the field of hybrid quantum system for future quantum information processing\cite{reentrant,Tabuchi:2014aa,Andrich:2017aa,Tabuchi:2015aa,Zhang:2014aa,Lachance-Quirion:2017aa}. Also, it has been proposed to use a microwave cavity to read out axion induced magnetisation. An excited ferromagnetic sample in free space will dissipate energy by dipole radiation. Placing it in a microwave cavity, therefore, has the additional benefit of mitigating this dissipation by collecting the radiation in the cavity. Thus, the haloscope based on axion-electron coupling appears as the hybrid mode cavity-ferromagnetic quasi particle (or cavity-magnon polariton) widely considered for applications in quantum information processing \cite{magnonDataProcessing,magnonSpintronics}.\\

To model the dynamics of a ferromagnet, it is common to use the Bloch equations to determine the induced magnetization from the pseudo-magnetic field. In the context of the QUAX proposal\cite{quaxprop,Crescini2018}, the solutions to these equations were modified in conjunction with the cavity equations of motion in the fully hybridised regime assuming the magneto-static and cavity mode form coupled harmonic oscillators. Analysis of this system, however, needs to be extended to the dispersive regime. To model the system in a single framework, this work considers a Hamiltonian approach. The Hamiltonian of the total detector system consists of its cavity and magnon parts, $H_c$ and $H_m$ respectively, as well magnon-cavity and axion-magnon interaction parts, $H_\textrm{int}$ and $H_{aee}$ respectively (for units where $\hbar=1$):
\begin{multline}
\begin{aligned}
H &= H_c + H_m + H_\textrm{int} + H_{aee} \\
H &= \omega_c c^\dagger c + \omega_m b^\dagger b +  g_{cm}(c^\dagger+ c)(b^\dagger + b) + H_{aee},
\end{aligned}
\end{multline}
where $c^\dagger$ ($c$) is a creation (annihilation) operator for photon, $b^\dagger$ ($b$) is a magnon creation (annihilation) operator, $\omega_c$ is the cavity frequency, $\omega_m$ is magnon frequency, and $g_{cm}$ is the cavity-magnon coupling rate. These expressions can been found from first principles \cite{cavitymagnon, magnonkerr}. It can also be noted that the magnon Hamiltonian assumes the demagnetizing and magneto-crystalline anisotropic field do not introduce non-linearities into the dynamics. These assumptions are satisfied using a spherical sample and small input fields respectively\cite{magnonkerr}. The magnon resonance frequency is set by the external magnetic field $B_0$ via the Zeeman effect $\omega_m=\gamma (B_0+B_\textrm{ani})$, where $\gamma$ is the gyromagnetic ratio and $B_\textrm{ani}$ is the linear contribution of the magneto-crystalline anisotropy, typically determined experimentally.\\

To determine the magnon-axion interaction term in the Hamiltonian, it is straightforward to write the interaction in terms of its Zeeman energy:
\begin{align}
H_{aee} &= -\int_{V_m}\mathbf{M}\cdot\mathbf{B}_{aee}\mathrm{d}V,
\end{align}
where $\mathbf{M}$ is the sample magnetization vector, $V_m$ is its volume. 
Given the uniform precession magneto-static (Kittel) mode, which has constant magnetisation over the sample volume, the expression can be easily integrated:
\begin{align}
H_{aee} =& -\gamma \mathbf{S}\cdot\mathbf{B}_{aee},
\end{align}
where $\mathbf{S}=\frac{\mathbf{M}V_m}{\gamma}$ is a macrospin operator. For this operator, raising and lowing spin operators are introduced as $S^{\pm}=S_x\pm iS_y$, where for the following, the $z$ direction is chosen to align with the external DC magnetic field. 
Using this set of spin operators, the spin-axion interaction Hamiltonian is expanded as:
\begin{align}
H_{aee} =& -\gamma B_{aee}\Big[\frac{\sin(\phi)}{2}(S^+e^{-i\theta} + S^-e^{i\theta}) + S_z \cos(\phi)\Big],
\end{align}
where $\theta$ and $\phi$ are the spherical coordinate azimuthal and polar angles of the pseudo-magnetic field vector respectively.  Finally, the Primakoff-Holstein transformations are used to write the interaction it in terms of magnon operators introduced above \cite{PhysRev.58.1098}:
\begin{multline}
\begin{aligned}
S^+ = (\sqrt{2S - b^\dagger b})b,\\
S^- = b^\dagger(\sqrt{2S - b^\dagger b}),\\
S_z = S - b^\dagger b,
\end{aligned}
\end{multline}
where $S$ is the total spin number of the macrospin operator. This number is determined by $S=\frac{\mu}{g\mu_B}N_s$, where $\mu$ is the magnetic moment of the magnetic sample, $\mu_B$ is the Bohr magneton, $g$ is the g-factor ($g=2$) and $N_s$ is the number of spins in the sample (given by $N_s=n_sV_m$ with $n_s$ as spin density and $V_m$ as volume). For YIG, used in previous proposals as well as this work, it is estimated that $\frac{\mu}{\mu_B}=5.0$\cite{blundell} and the spin density
$n_s=2.2\times10^{28}$m$^{-3}$. From these values the total spin number for a 2mm diameter YIG sphere used in this work may be estimated as $S=2.3\times10^{20}$.\\

For low excitation numbers (${\langle b^\dagger b\rangle}\ll 2S$), the macrospin operators may be approximated by $S^+\approx\sqrt{2S}b$ and $S^-\approx\sqrt{2S}b^\dagger$ giving the final expression for the axion-magnon coupling term:
\begin{multline}
\begin{aligned}
H_{aee} = -\gamma B_{aee}\Big[\sqrt{\frac{S}{2}}\sin(\phi)(be^{-i\theta} + b^\dagger e^{i\theta}) +\\ b^\dagger b \cos(\phi)\Big],
\label{final}
\end{aligned}
\end{multline}
where constant terms are ignored. It can be noted the axion-magnon interaction Hamiltonian (\ref{final}) includes two sets of terms: components relating to the pseudo-magnetic field perpendicular to the external magnetic field and producing magnon excitations, and terms coupled to the parallel component producing a modulation of the magnon frequency. For an axion producing magnons in the GHz frequency range, it is shown in the following discussion that the latter frequency modulation term can be ignored under a Rotating Wave Approximation (RWA), but could be used to search for low mass axions using frequency metrology techniques \cite{axionMetrology,exceptionalAxions}. Thus, only the former term producing excess of power in the magnon spectrum can be used for the current experiment. 

\subsection{Detector Dynamics}

In order to capture the dynamics of the total experiment, one needs to include dissipation. This is done  by deriving the Heisenberg-Langevin equations of motion for open systems. In this case internal dissipation in the cavity and magnon modes are modelled as coupling to internal thermal baths as well as the cavity coupling to an external bath through a probe/port. The general form of these equations is given by \cite{qoptics}:
\begin{multline}
\begin{aligned}
\deriv{f} = -i[f,H] + \sum_{n}\Big(\Big(\frac{\kappa_n}{2}g^\dagger+i\sqrt{\kappa}_n a_n^{\textrm{in}\dagger}\Big)[f,g]-\\
[f,g^\dagger]\Big(\frac{\kappa_n}{2}g+i\sqrt{\kappa}_n a_n^\textrm{in}\Big)\Big),
\end{aligned}
\end{multline}
where $f$ and $g$ are arbitrary system operators, $a_n^\textrm{in}$ and $\kappa_n$ is the input field and dissipation rate due to coupling of some $a$ field to the $n$th thermal bath respectively. For derivation of axion induced output power, all noise terms are ignored, and it is assumed that the magnon mode does not couple to the external excitation port.

 The resulting equations of motion for annihilation operators are then converted to the rotating frame ($c\rightarrow ce^{-i\omega_a t}, b \rightarrow be^{-i\omega_a t}$, where $\omega_a$ is the axion signal frequency) and a RWA is made. This procedure gives the equations of motion as follows in the lab frame:
\begin{align}\label{EOM}
\deriv{c} &= -i\Big(\omega_c-i\frac{\kappa_c}{2}\Big)c-ig_{cm}b,\\
\deriv{b} &= -i\Big(\omega_m-i\frac{\kappa_m}{2}\Big)b-ig_{cm}c +i \sqrt{\frac{S}{2}} \gamma B_{aee}\sin(\phi)e^{i\theta},
\end{align}
where $\kappa_c=\kappa_c^\textrm{int}+\kappa_c^\textrm{ext}$ and $\kappa_m=\kappa_m^\textrm{int}$ are the dissipation rates of the cavity and magnon respectively. These quantities include internal and external (from the cavity probe) dissipation where applicable. 

The derived equations of motion (\ref{EOM}) can be easily solved for the cavity field in terms of the driving pseudo-magnetic field giving:
\begin{equation}
c(\omega_a) = \frac{ig_{cm}\sqrt{\frac{S}{8}} \gamma B_{aee}\sin(\phi)e^{i\theta}}{\kappa_c\kappa_m\Delta_c\Delta_m-g_{cm}^2},
\end{equation}
where $\Delta_n = \frac{\omega_a-\omega_n}{\kappa_n}+\frac{i}{2}$. This result can be converted to the output fields using the boundary conditions from input-output theory $c_\textrm{out}+c_\textrm{in}=\sqrt{\kappa_c^\textrm{ext}}c$, as well as the output power from $P_\textrm{out}=\hbar \omega_a \langle c^*_\textrm{out}(t)c_\textrm{out}(t)\rangle$:
\begin{equation}
P_\textrm{out}(\omega_a) = \frac{\hbar\omega_a\kappa_c^\textrm{ext}g_{cm}^2\frac{S}{8} \gamma^2 B_{aee}^2\sin^2(\phi)}{|\kappa_c\kappa_m\Delta_c\Delta_m-g_{cm}^2|^2}.
\end{equation}
Note should be taken that the only geometrical dependence left in the output power is $\sin^2(\phi)$, which is maximum when the pseudo-magnetic field from axion interaction is perpendicular to the DC magnetic field. This dependence is discussed in Section~\ref{windS}.\\

\begin{figure}[h]
	\includegraphics[width=\columnwidth]{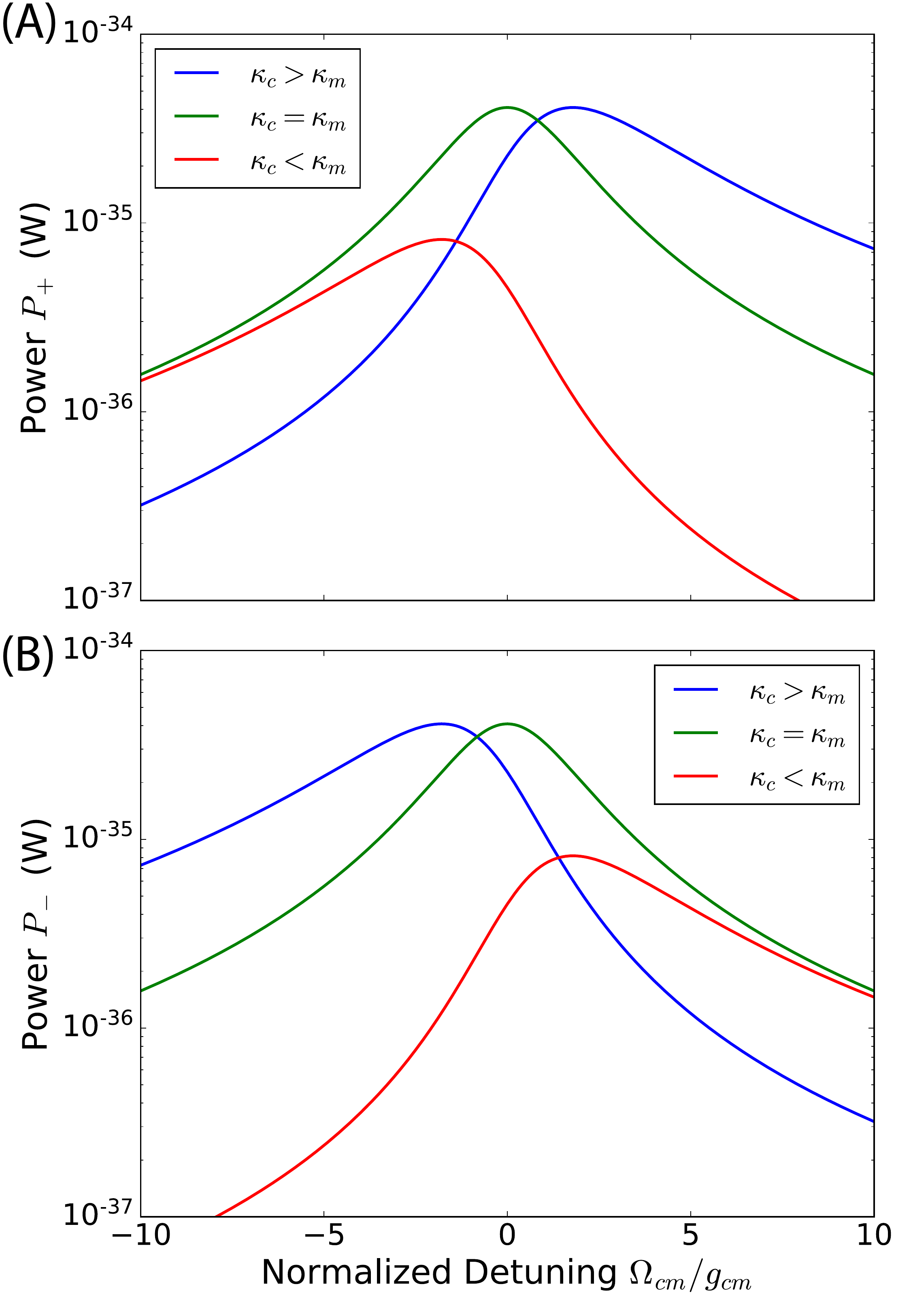}
	\caption{Detector output power as a function of scaled detuning $\frac{\Omega_{cm}}{g_{cm}}$ for an YIG sphere for various linewidths: (A) from the higher frequency normal mode ($P_+$), (B) from the lower frequency normal mode ($P_-$).}
	\label{power}
\end{figure}

To determine the axion induced power on resonance, the resonant frequency of the coupled normal modes can be used, where a simplification has been made under the assumption of a strongly coupled system ($g_{cm} \gg \kappa_c,\kappa_m$):
\begin{equation}\label{hybridFs}
\omega_\pm = \frac{\omega_c+\omega_m}{2} \pm \sqrt{\Big(\frac{\Omega_{cm}}{2}\Big)^2+g_{cm}^2},
\end{equation}
where $\Omega_{cm} = \omega_m - \omega_c$. These expression give the on resonance ($\omega_a=\omega_\pm$) power expressions:
\begin{equation} \label{FOM}
P_{\pm} = \frac{\hbar\omega_a\kappa_c^\textrm{ext}\frac{S}{2} \gamma^2 B_{aee}^2\sin^2(\phi)}{\frac{\kappa_c^2\kappa_m^2}{4g_{cm}^2} + \Big(\frac{\Omega_{cm}}{2g_{cm}}(\kappa_m-\kappa_c) \pm \sqrt{\Big(\frac{\Omega_{cm}}{2g_{cm}}\Big)^2+1}(\kappa_c+\kappa_m)\Big)^2}.
\end{equation} 
It can be noted that this form of the power on resonance is a function of the detuning over coupling rate. Thus, the range of frequencies where this experiment is sensitive is proportional to the coupling rate $g_{cm}$. Importantly, this formula allows the determination of the sensitivity of this experiment in the dispersive regime when $\omega_c\ne\omega_m$. To demonstrate the response of the system in the dispersive regime a set of output powers are calculated for a 2mm diameter YIG sphere for various relationships between linewidths of subsystems. Given linewidths $\kappa_1=(2\pi)10$MHz and $\kappa_2=(2\pi)50$MHz, three cases are considered: (1) for $\kappa_c>\kappa_m$, $\kappa_c=\kappa_1$ and $\kappa_m=\kappa_2$, (2) for $\kappa_c=\kappa_m=\kappa_1$, (3) for $\kappa_c<\kappa_m$, $\kappa_c=\kappa_2$ and $\kappa_m=\kappa_1$. In all cases $\kappa_c^\textrm{ext}=\kappa_c/2$. The result  for $g_{cm} = (2\pi)745$MHz, $\omega_a = (2\pi)8.2$GHz (corresponding to $|B_{aee}|=3.4\times10^{-23}$T) and $\sin(\phi) = 1$ is depicted in Fig.~\ref{power}. As the magnon excitation is readout through the cavity mode, this might imply that the output signal will be strongest in the full hybridisation regime. However, Fig.~\ref{power} shows that this is not the case in general. Particularly, when $\kappa_c\neq\kappa_m$, the maximum signal strength is obtained in the dispersive regime. This opens up new ways to optimise such an experiment, often by operating in the dispersive regime.\\

\section{Axion Wind}
\label{windS}

With the detector dynamics determined in terms of the driving pseudo-magnetic field, it is now necessary to determine its components in the laboratory frame. As such a classical axion field can be substituted (a step which will be justified later). This is given by:
\begin{equation}
a(\mathbf{r},t) = a_0\sin\Big(\omega_at-\frac{m_a \mathbf{v_a} {\cdot} \mathbf{r}}{\hbar}\Big),
\end{equation} 
where $a_0 = \sqrt{\frac{2\rho_a}{c}}\frac{\hbar}{m_a}$ is the magnitude of the axion field, $\rho_a\approx0.45$GeV/cm$^3$ is the local dark matter density\cite{DMdensity}, and $\mathbf{v_a}$ is the velocity of the axion wind.
\begin{equation}
\mathbf{B}_{aee} = \frac{g_{aee}}{2e}\frac{m_a |\mathbf{v_a}|}{\hbar}a_0\sin\Big(\omega_at-\frac{m_a |\mathbf{v_a}| x}{\hbar}\Big)\mathbf{\hat{x}},
\label{wind}
\end{equation}
where $\mathbf{\hat{x}}$ is the direction of the axion wind. In the present work, the velocity in the galactic frame of the axion wind is treated as a Maxwell distribution with mean velocity of zero and velocity dispersion of $\sigma_v=270$km/s. Thus, in the laboratory frame it has the velocity of the galactic frame of approximately $220$km/s in the direction of the constellation Cygnus \cite{dispersion}. It should be noted that this experiment being a directional detector of the axion dark matter field makes it also sensitive to the so called `dark matter hurricane' or S1 stream which provides an additional source of axions as well as different velocity distribution compared with those in the dark matter halo \cite{DMhurricane}. In this work, however, we only consider the axion wind from the earth's movement through the dark matter halo which gives the following parameters\cite{Crescini2018}:
\begin{align}
|B_{aee}| &= 3.4\times 10^{-23} \Big(\frac{m_a}{34\mu \mathrm{eV}}\Big)\mathrm{T}, \\
\frac{\omega_a}{2\pi} &= 8.2\Big(\frac{m_a}{34\mu \mathrm{eV}}\Big)\mathrm{GHz},\\
g_{aee} &= 1.0\times 10^{-15}\Big(\frac{m_a}{34\mu \mathrm{eV}}\Big),\\
\lambda_{\nabla a} &= 0.74\lambda_a=0.75\frac{h}{m_av_a}=8.8\Big(\frac{34\mu \mathrm{eV}}{m_a}\Big)\mathrm{m}.
\end{align}
Note the de-Broglie wavelength of galactic axions is much larger than the scale of a haloscope experiment for the frequencies considered.
This justifies the treatment of the axion field $a(\mathbf{r},t)$ as a classical field.\\

\begin{figure}[h]
	\centering
	\includegraphics[width=\columnwidth]{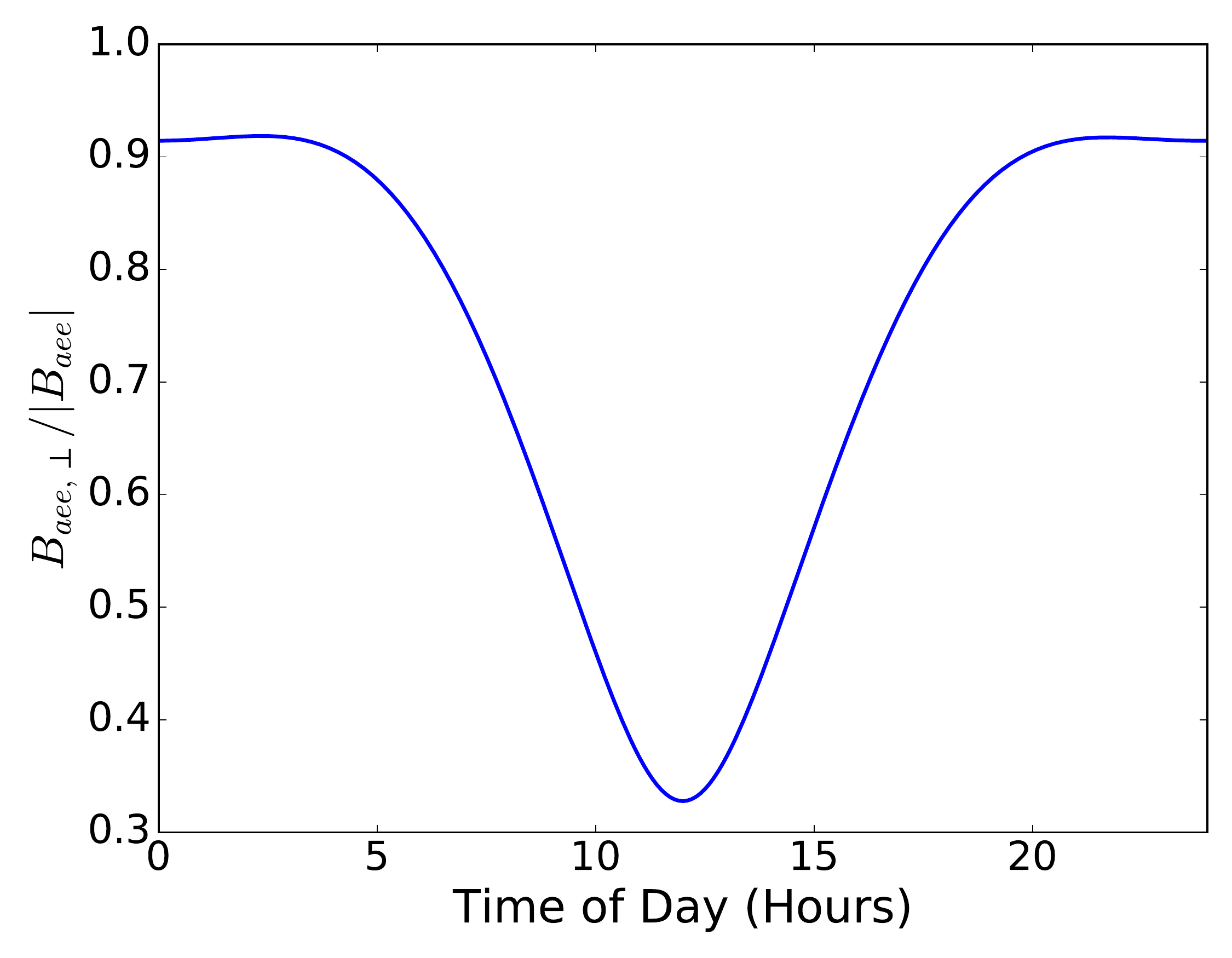}
	\caption{The daily modulation of the pseudo-magnetic field perpendicular to the direction of an external field at the University of Western Australia on the 27th of August 2018.}
	\label{awind}
\end{figure}

The next step is to determine the direction of the axion wind with the laboratory frame in Perth, Western Australia. For this, the velocity due to the Sun's orbit around the galactic centre as well as the velocity due to the Earth's orbit around the Sun can be taken into account. Appropriate coordinate transforms \cite{celestialcoords} give the velocity of the axion wind in the laboratory frame. In these calculations, only the components which are perpendicular to the external DC magnetic field oriented locally upwards (from the centre of the Earth), are important to the operation of the haloscope. The magnitude of the perpendicular components of the velocity $B_{aee,\perp}=\sqrt{B_{aee,x}^2+B_{aee,y}^2}$ is shown in Fig.~\ref{awind} for the 27th of August 2018 (local time) in Perth where it is scaled to its absolute value. It can be seen that there is an $~8$ hour period per day which a ferromagnetic haloscope would be sensitive and a $~2$ hour period where our signal should disappear. 

\section{Experimental design}

From the expression of output power, Eq.~(\ref{FOM}), it follows that in order to maximise the detector sensitivity it is required to maximise the number of spins (i.e. spin density and volume) as well as cavity and magnon mode Quality factors. For a scanning tunable experiment, it is also important to widen the range of sensitive frequencies that can be done by improving the coupling rate between cavity and magnon modes. This has been achieved using two post re-entrant cavities allowing to reach the strong coupling regime\cite{reentrant}. As a system of essentially two coupled quasi-lumped resonators, such cavities exhibit two modes: the bright and the dark modes.
In the former mode, the magnetic field around each post as well as the electric fields in the gaps are in phase. This phase relation concentrates the magnetic energy in between the posts. So, by placing the magnetic sample in that space, one achieves large magnetic filling factors and thus a large coupling between the photon and magnon mode. \\

\begin{figure}[h]
	\hspace{-0.9\columnwidth}
	\parbox[b][3.5cm][t]{1em}{(A)}
	\hspace{0.1\columnwidth}
	\parbox[c]{0em}{\includegraphics[width=0.8\columnwidth]{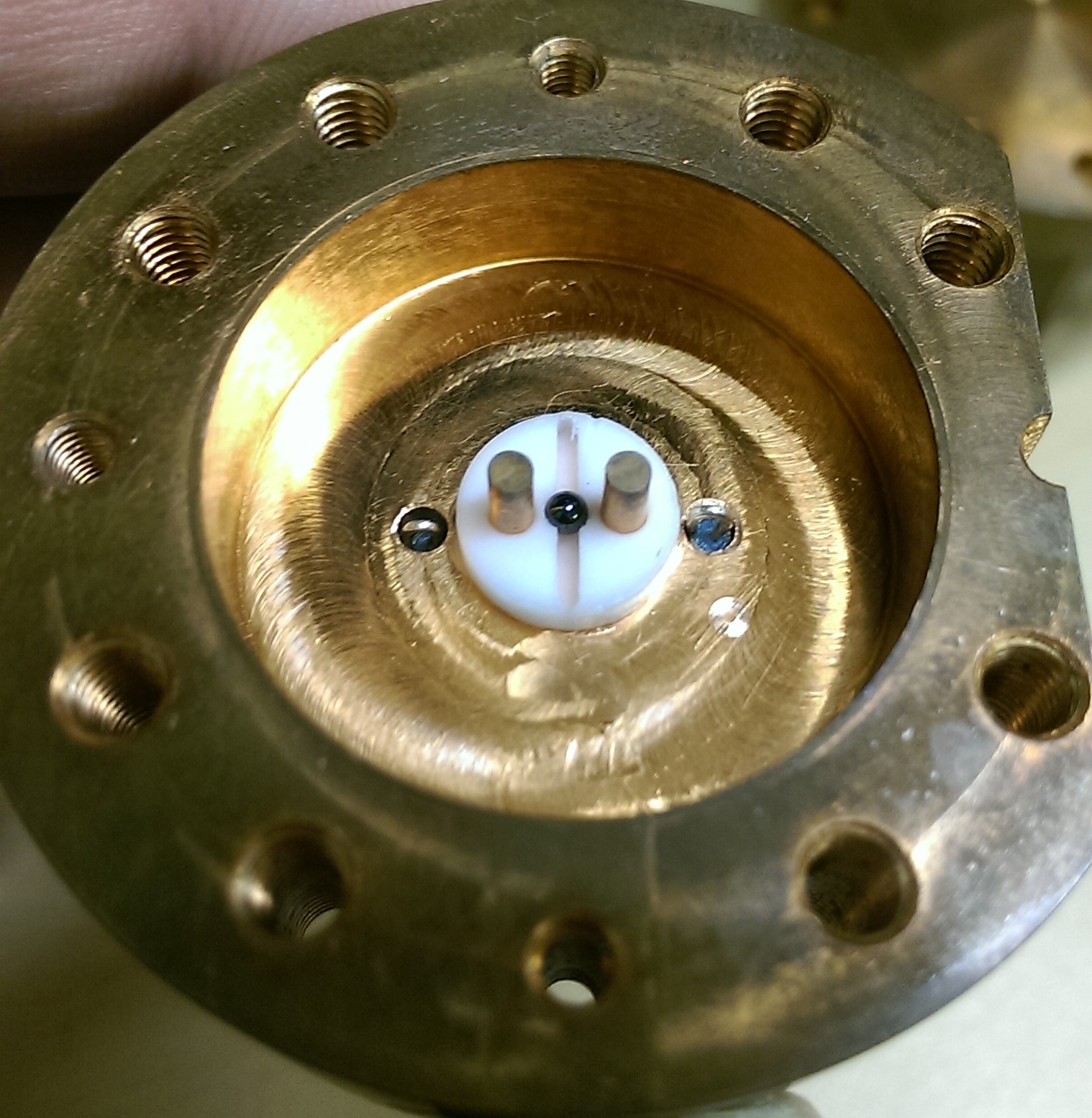}}\\
	\hspace{-0.95\columnwidth}
	\parbox[b][2.8cm][t]{1em}{(B)}
	\hspace{0.05\columnwidth}
	\parbox[c]{0em}{\includegraphics[width=0.9\columnwidth]{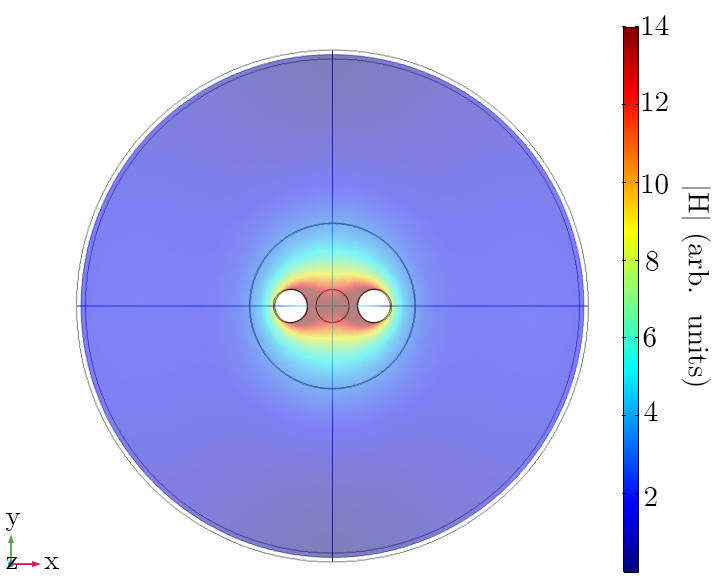}}
	\caption{(A) A picture of the detector cavity without its lid. Visible is the loop probes, the re-entrant cavity posts, the Teflon holder and the YIG sphere. (B) Colour density plot of magnetic field strength $|\mathbf{H}|$ for the bright mode of the two post re-entrant cavity. }
	\label{system1}
\end{figure}

To achieve large spin density and magnon quality factors, a 2mm diameter single domain crystal YIG sphere is used. The sphere is placed in a Teflon holder at the centre of the copper two post re-entrant cavity. This cavity is a 30$\times$8mm (diameter $\times$ height) cylinder with two identical posts of 2$\times$7mm at a separation of 2.5mm from the cavity centre. A photo of this assembly is shown in Fig.~\ref{system1} (A). Finite Element Modelling (FEM) was used to determine the cavity resonance frequencies and the mode shapes which for the bright mode is shown in Fig.~\ref{system1} (B). 

\begin{figure}[h]
	\hspace{-0.75\columnwidth}
	\parbox[b][3.6cm][t]{0em}{(A)}
	\hspace{0.25\columnwidth}
	\parbox[c]{0em}{\includegraphics[width=0.5\columnwidth]{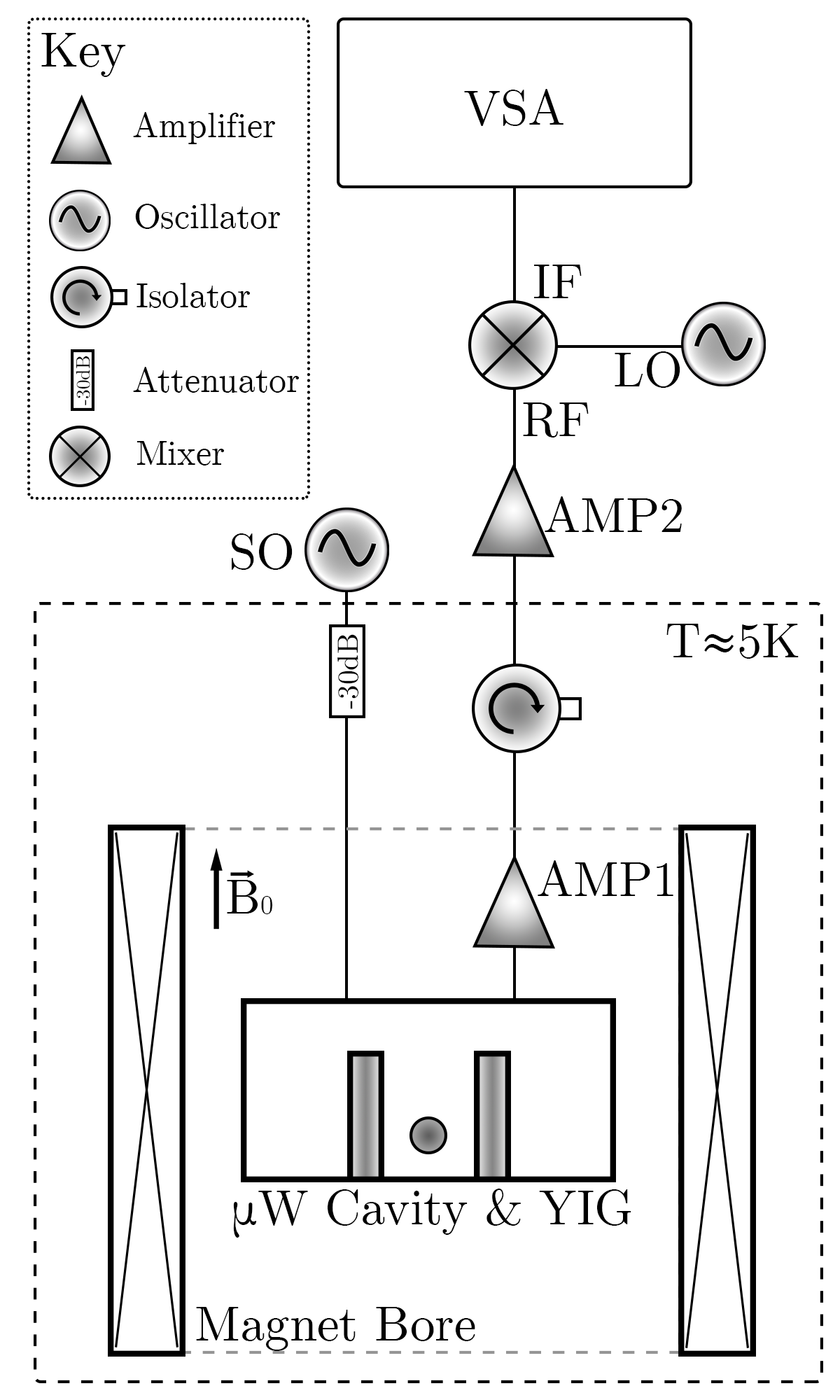}}\\
	\hspace{-\columnwidth}
	\parbox[b][2.5cm][t]{2em}{(B)}
	\parbox[c]{0em}{\includegraphics[width=\columnwidth-2em]{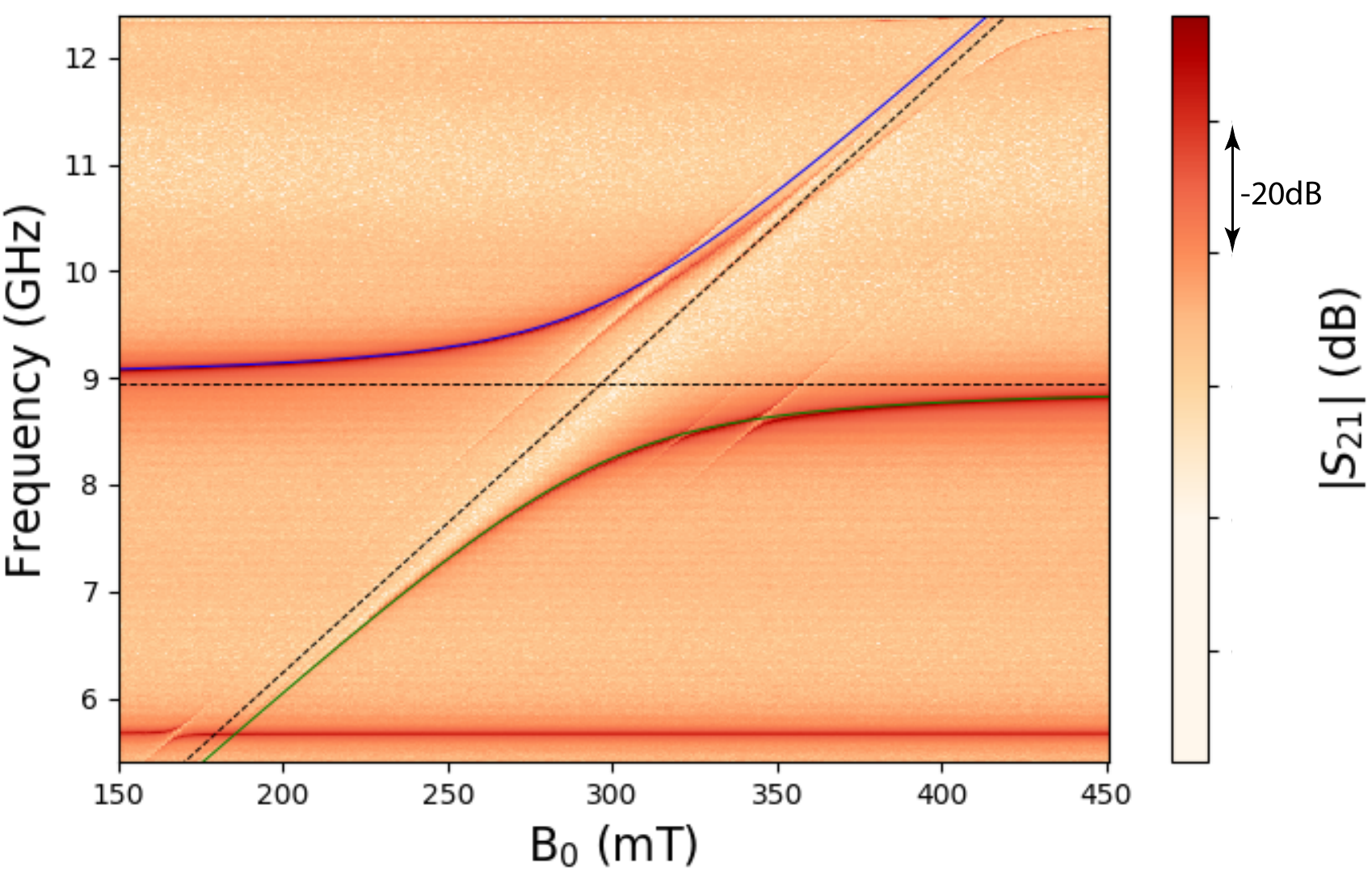}}
	\caption{(A) Experimental setup used to characterise the detector (involving SO source) and final data acquisition. (B) System response in terms of $S$-parameters as a function of external magnetic field.}
	\label{setup}
\end{figure}

To characterise the detector in terms of losses and couplings, the system response in terms of scattering parameters is measured as a function of external magnetic field. A simplified schematic of the corresponding experimental setup is sketched in Fig.~\ref{setup} (A). The cavity and sphere are placed inside the bore of a superconducting magnet held at $T\approx5$K. A source oscillator (SO) is attenuated and fed to a weakly coupled probe in the cavity for spectroscopic measurements. The readout chain consists of a cryogenic High Electron Mobility Transistor (HEMT) amplifier based low noise amplifier bolted directly to the cavity (AMP1), an isolator to prevent reflected noise, a room temperature low noise amplifier (AMP2) fed into the RF port of a microwave mixer. The signal is then mixed with a Local Oscillator (LO) downconverting the signal to a few MHz and finally measured by a vector signal analyser (VSA). AMP1 has a gain of $41.5$~dB and a noise temperature of $~4$K at $8.2$~GHz quoted by the manufacturer, and AMP2 had a measured gain of $21$dB. For the initial spectroscopic measurements, the SO, mixer, the LO and the VSA are substituted by a Vector Network Analyser that measures the corresponding $S$-parameters.
The resulting spectroscopic data alongside with two mode system fit are shown in Fig.~\ref{setup} (B).\\

From Fig.~\ref{setup} (B), the cavity frequency is estimated as $f_c=8.937$~GHz, the magnon frequency as a function of external magnetic field could be approximated by $f_m(B_0)=27.99B_0$(GHz/T)$+0.660$(GHz), and the magnon-photon coupling rate is $g_{cm}=751$~MHz. 
The uncoupled linewidths can also be inferred from spectroscopic data giving $\kappa_m=(2\pi)11.1$~MHz and $\kappa_c=(2\pi)7.53$~MHz (from fully hybridised values $\kappa_\pm=(2\pi)9.32$~MHz where in general $\kappa_c+\kappa_m = \kappa_++\kappa_-$). The input probe of the cavity is only weakly coupled for transmission measurements which can be considered a negligible source of loss. Prior to running this experiment reflection measurements of the output cavity probe were performed to infer the level of coupling to the photon mode as $\kappa_c^\textrm{ext}=(2\pi)5.48$MHz. The output probe needs to be near critically coupled to maximise power output. The measured parameter $\kappa_c^\textrm{ext}$, implies coupling to the hybrid mode when fully hybridised as $\beta = \frac{\kappa_c^\textrm{ext}}{\kappa_\pm - \kappa_c^\textrm{ext}} \approx 1.4$, where $\beta=1$ is the condition for critical coupling.\\

\section{Results}

The axion signal is expected to appear as excess of power over background noise in a power spectra. It is therefore necessary to discriminate against artificial peaks that could appear due to various spurious signals. For this reason, two sets of data for each frequency range are taken each day. The first is over an 8hr period where the velocity is $\sim90\%$ perpendicular to the external magnetic field (corresponding to $\sim81\%$ of the maximum power) and then a second set of data is taken over the 2 hours period when the velocity is only $\sim35\%$ perpendicular to the external field (corresponding to $\sim12\%$ of the maximum power). The second dataset allows any persistent spurious signals to be identified and eliminated. \\

The exclusion data was taken over a range of frequencies for a period of 6 days in August 2018. Each day of measurements consists of 320,000 individual power spectra measured over a sensitive period of time ($\sim8$hrs integration time) and 80,000 power spectra collected over the minimally sensitive period ($\sim2$hrs integration time). The frequency of the hybrid resonance was tuned using the external magnetic field allowing to scan for $\sim36$MHz around $8.2$GHz. In this case only the lower hybrid mode was used to  attempt to measure an axion signal and the a limited $36$MHz tuning range used considered due to availability of equipment. The full potential range of this haloscope can be calculated from equations \ref{hybridFs} and \ref{FOM} giving a full width at half maximum (FWHM) of $1.6$GHz around $10$GHz for the upper hybrid mode and a FWHM of $1.0$GHz around $8.2$GHz for the lower hybrid mode. Throughout the experiment the cavity and first amplifier were kept at a temperature of $\sim5$K. The sum of the cavity physical temperature and the noise temperature of the first stage amplifier can be used to estimate the expected noise temperature of the system as $\sim9$K at $8.2$GHz. Analysis of the spectra was done using a similar method to E. Daw \cite{daw}. For each set of data, defined by the DC magnetic field $B_0$ used to set a central frequency of the lower hybrid mode, $f_-$, a span of $\sim6$MHz (reduced from the $\sim9$MHz hybrid linewidth due to spurious system noise) of data around the central frequency was analysed. In each case, the power spectral density, with bin width $\Delta f=3.125$kHz (chosen to be similar to the axion signal width $4.1$kHz), was scaled by the amplifier gains and a polynomial fit was made. The residuals were then analysed to confirm the validity of the fit. Next the lowest possible cut is made in the sensitive data such that any bins above this chosen power value also exist above the same cut in the insensitive data. This allows the identification of bad bins, as any bins above the cut in both the sensitive and insensitive data are irrelevant spurious signals and thus can be removed from the analysis. The remaining Gaussian noise can be analysed to determine standard deviation ($\sigma$) and effective noise temperature of the system, making use of the Dicke radiometer equation \cite{radiometer}:
\begin{equation}
\sigma = k_BT_\textrm{eff}\sqrt{\frac{\Delta f}{t}},
\end{equation}
where $k_B$ is the Boltzmann constant, $T_\textrm{eff}$ is the effective system noise temperature, and $t$ is the integration time. A potential axion signal would appear as a residual bin with an excess of power over the mean, with the cut made earlier defining the largest distinguishable signal amongst the noise. Gaussian statistics are then used to determine the excluded signal power ($P_\textrm{exl}$) to a 95\% confidence based on the cut made \cite{daw}. These results are shown in Table~\ref{table1}. From Table~\ref{table1}, it can be seen that the effective temperature determined from the measured noise is comparable to the estimate based on the physical temperature and first stage amplifier, thus the measured results are consistent with expectations. The larger measured effective noise temperature is due to the small contribution of noise due to cables and second stage amplifier.\\

\begin{table}[]
	\begin{tabular}{| l | l | l | l | l | l |}
		\hline
		$B_0$ (mT)   & $f_-$ (GHz) & $\sigma$ ($10^{-23}$W) & $T_\textrm{eff}$ (K)   & Cut ($\sigma$) & $P_{exl}$ ($\sigma$)  \\\hline
		298          & 8.2086     & 4.89  & 11.0     & 6.2 & 15.8  \\\hline
		297.5        & 8.2018     & 4.73  & 10.4     & 4.2 & 10.3  \\\hline
		297      	 & 8.1950     & 5.81  & 12.8     & 5.0 & 13.2  \\\hline
		296.5        & 8.1879     & 5.50  & 12.1     & 5.7 & 14.9  \\\hline
		296.1        & 8.1820     & 5.74  & 12.6     & 5.0 & 12.9  \\\hline
		295.6        & 8.1754     & 5.78  & 12.7     & 5.2 & 13.6  \\\hline
	\end{tabular}
\caption{Measured standard deviations of residuals (scaled by amplifier gains), cuts made and excluded power for each set of data.}
\label{table1}
\end{table}

\begin{figure}[h]
	\centering
	\includegraphics[width=\columnwidth]{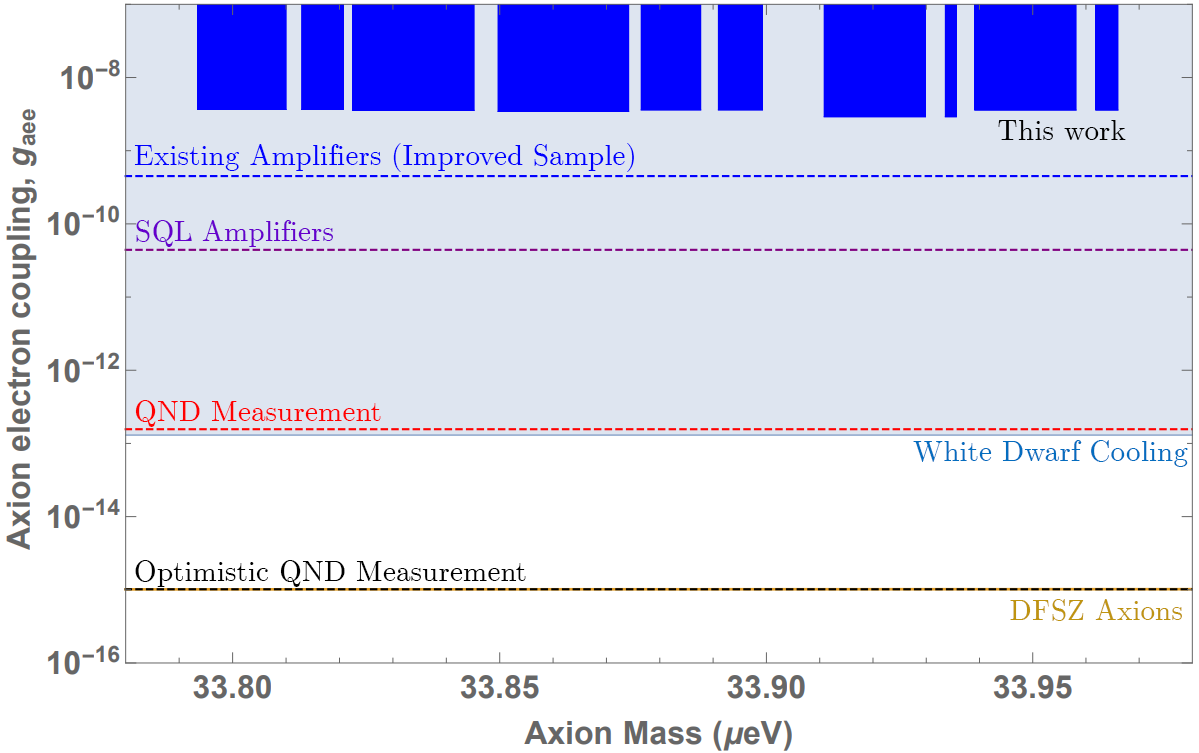}
	\caption{The DFZS axion model band and exclusion plot for axion-electron coupling strength $g_{aee}$ as a function of axion mass: limits due to white dwarf cooling \cite{whitedwarf1, whitedwarf2} are in light blue, this work limits are in dark blue, dashed lines show several predictions for the future work.}
	\label{exclusion}
\end{figure}

To determine excluded couplings and magnetic fields the following relations were used:
\begin{equation}\frac{P_\textrm{exl}}{P_{-}}=\frac{g_{aee,\textrm{exl}}^2}{g_{aee}^2},
\end{equation}
where $g_{aee,\textrm{exl}}$ is the excluded axion to electron coupling strength. Here the excluded power is also scaled by a Lorentzian line-shape and multiplied by $0.81$ to account for the reduction in sensitivity due to the proportion of axion wind perpendicular to the external magnetic field at that time of day. This relation allows to put limits on the axion-electron coupling strength $g_{aee}$ as shown in Fig.~\ref{exclusion} where also several predictions are made in the form of the dashed lines (see Section~\ref{disc} for detailed discussions).

\section{Discussion and Perspectives}
\label{disc}
It can be seen in Fig.~\ref{exclusion} that the results of this analysis are still orders of magnitude from current astronomical limits on axion electron coupling and expected DFSZ model predictions. They do, however, demonstrate how such ferromagnetic haloscopes, in a single cavity configuration, can search over a larger range than their linewidth by tuning their hybrid frequencies. This larger range isn't strictly experimental improvements but rather from improved capabilities of the experiment due to more general theoretical analysis. Additionally, whilst this detector isn't sensitive enough to detect DFSZ axions, it does have the capability to detect axion-like particles (ALPs) which don't have a fixed relation between axion mass and axion-normal matter coupling strengths.\\

While this initial result demonstrates the experiments usability, further improvements, particularly in sample volume and line-widths, are needed to improve the overall sensitivity. Single domain ferrimagnetic spheres with a diameter of 5mm are easily available and inferred magnon linewidths of around $2.4$MHz at central frequency $14$GHz can be seen for a Ga:YIG sample in the QUAX experiment \cite{Crescini2018}. Additionally, their approach to increasing the effective volume of the magnetic material in the cavity by including multiple spheres could also be utilised. Such improvements would boost the output power from an axion signal. The corresponding improvement is shown in Fig.~\ref{exclusion} by the dark blue dashed line assuming a larger 5mm diameter sphere, improved linewidths of $\kappa_m=(2\pi) 2.4$MHz and a measurement time of one day with existing amplifiers. Alternatively, one may consider materials with higher spin density such as Lithium Ferrite where the same order of magnetic losses together with absence of spurious modes have been observed\cite{Goryachev:2018aa}.\\

The sensitivity would also be improved by reducing the detector background noise, for example, by implementation of quantum limited parametric amplifiers based on Josephson junctions. Successful operation of such amplifiers, viable in the considered frequency range, has been demonstrated for axion-photon haloscopes\cite{haystac}. The result of these improvements changing sensitivity of the detector is shown in Fig.~\ref{exclusion} with the purple dashed line.\\

In order to improve the limits even further, one needs to build a viable detection scheme based on a single photon or magnon counter, ideally using a Quantum Non-Demolition (QND) measurement to surpass the standard quantum limit. Such devices would then only be limited by shot noise due to a non-zero temperature of the cavity. Single photon counters in the context on axion haloscopes and are discussed by Lamoreaux, et al.\cite{QNDforhaloscopes}, showing superconducting qubits in cavities to be a promising avenue for a QND single photon counter \cite{QNDsuperconducting}. A single magnon counter could similarly be constructed by coupling a qubit to a magnon mode which was recently achieved to resolve numbers state \cite{magnonMeetsQubit,resolvingmagnons}. This is one experimental advantage of ferromagnetic haloscopes over traditional photon haloscopes as the DC magnetic field need not extend over the entire cavity, thus a focused magnetic field on the ferrimagnetic sample would allow the presence of superconducting devices in the cavity to aid measurement. Non-QND, single photon counters have also been shown to be another promising avenue for axion haloscopes \cite{nonQNDphotoncounter, coldeletron}.  The result of a QND measurement based experiment is predicted in Fig.~\ref{exclusion} with both reasonable and optimistic improvements in the signal-to-noise ratio. The red dashed line is the result of the above assuming a perfect efficiency QND measurement is achieved requiring a maximum physical temperature of $12.5$mK to ensure a 95\% confidence of no dark counts of the detector over the measurement time. The final black dashed line in Fig.~\ref{exclusion} is a prediction of an extremely optimistic QND measurement scheme assuming the signal power can be further boosted with a volume of the magnetic sample of $V_m = 0.13$~L and an improved linewidth of $\kappa_m=(2\pi)200$~kHz. These estimations are done following the procedure by Lamoreaux et al. \cite{QNDforhaloscopes}, where it is noted that in this case the limiting factor is the minimum detectable power required to count at least three photons over the measurement time.\\

\section*{Conclusions}
A new theoretical perspective on ferromagnetic haloscopes was presented which demonstrated methods of improving the operation of a ferromagnetic haloscope. Particularly easy frequency tunability to search the axion mass parameter space is achievable by operating in the dispersive regime. It also highlighted the importance of a large cavity-magnon coupling strength to produce a large bandwidth and provided new methods to optimise experimental design and operation. The device was implemented and set limits on axion-electron coupling of $g_{aee}>3.7\times10^{-9}$ in the range $33.79\mu$eV$<m_a<33.94\mu$eV with 95\% confidence. Limitations of this experiment, in the form of electronic noise and small sample volume are discussed, and improvements are suggested in the form of quantum or shot noise limited measurement. These predictions highlight the need for shot noise limited measurement in the form of single photon or magnon counters, as well as the requirement to dramatically improve signal power.

\section*{Acknowledgements}
This work was supported by the Australian Research Council grant number DP190100071 and CE170100009 as well as the Australian Government's Research Training Program.

\section*{References}
\bibliography{refs}

\end{document}